\documentclass[prl,aps,twocolumn,amsdtx,showpacs]{revtex4}
\usepackage{graphicx}
\usepackage{amsmath}
\usepackage{amssymb}
\usepackage[usenames]{color}

\newcommand{\bleq}{\ifpreprintsty
                   \else
                   \end{multicols}\vspace*{-3.5ex}{\tiny
                  \noindent\begin{tabular}[t]{c|}
                  \parbox{0.493\hsize}{~} \\ \hline \end{tabular}}
                   \fi}
\newcommand{\eleq}{\ifpreprintsty
                 \else
                   {\tiny\hspace*{\fill}\begin{tabular}[t]{|c}\hline
                    \parbox{0.49\hsize}{~} \\
                    \end{tabular}}\vspace*{-2.5ex}\begin{multicols}{2}
                    \fi}
\newcommand{\bcols}{\ifpreprintsty\else\begin{multicols}{2}\fi}
\newcommand{\ecols}{\ifpreprintsty\else\end{multicols}\fi}

\newcommand \beq  {\begin{equation}}
\newcommand \eeq  {\end{equation}}
\newcommand \bea {\begin{eqnarray} }
\newcommand \eea {\end{eqnarray}}

\begin{document}
\title{ A microscopic model for spiral ordering along (110) on the MnSi lattice}
\author{John M. Hopkinson}
\email{johnhop@physics.utoronto.ca}
\author{Hae-Young Kee}
\email{hykee@physics.utoronto.ca}
\affiliation{60 St.George St., University of Toronto, Toronto, Ontario, Canada}

\pacs{75.10.Hk,75.25.+z,71.10.Hf}

\date{\today}
\begin{abstract}
We study an extended Heisenberg model on the MnSi lattice.  In the cubic B20 crystal structure of MnSi, Mn atoms form lattices of of corner-shared equilateral triangles.  We find an ubiquitous spiral ordering along (110) for $J_1 <0$,
and $J_2=J_3 >0$, where $J_1, J_2,$ and $J_3$ are 1st, 2nd and 3rd nearest
neighbor Heisenberg interactions, respectively. While the ordering direction of (110) is reasonably robust to the presence of the Dzyaloshinskii-Moriya interaction, it can be shifted to the (111) direction  with the introduction of a magnetic anisotropy term for small $J_2/|J_1|$. We discuss the possible relevance of these results to the partially
ordered state recently reported in MnSi.

\end{abstract} 

\maketitle



{\it{Introduction:}} The helical magnetic ordering in MnSi at ambient pressures has been understood as a result of Dzyaloshinskii-Moriya (DM) interactions which arise due to the lack of inversion symmetry of the lattice{\cite{moriya}}.  Previous studies of symmetry respecting Landau-Ginzburg (LG) free energies have successfully described phenomena observed in MnSi{\cite{naganishi, bak,walker,Plumer}} including the wavevector reorientations induced by a magnetic field{\cite{walker}}. In the absence of an applied field, LG theories showed that ferromagnetic (FM) and DM terms lead to a helical ordering with a fixed periodicity, but no orientational preference.  The spiral ordering along (111) seen at ambient pressures can been understood as arising from crystal anisotropy terms{\cite{naganishi,Plumer}} consistent with the lattice symmetry which give energetic minima either at (111) or (100).  However, ordering along (110) cannot be obtained from an analysis of  LG
free energies solely consisting of the FM, DM, and
crystal anisotropy terms usually considered in MnSi.

It was thus of great interest when neutron scattering{\cite{pflei}} on MnSi under high pressures revealed unusual magnetic properties.  At pressures above $p_c=14.6kbar$, the neutron scattering intensity no longer sharply peaked along (111).  Instead, broad peaks at (110) with more diffuse scattering over a sphere of fixed magnitude wavevector have been reported in a small window of the pressure and temperature phase diagram.  The regime where this phenomenon has been observed has been dubbed the partially ordered state of MnSi. The magnitude of the ordering wavevector has not dramatically changed across $p_c$, but its orientation has clearly shifted from (111) to predominantly (110). It also remains to be understood why the resistivity has been found{\cite{nicholas1,Jaccard}} to be proportional to $T^{\frac{3}{2}}$ at and well above $p_c$, which was not anticipated by itinerant models{\cite{hertz-millis}}.  
\begin{figure}
\includegraphics[scale=0.4]{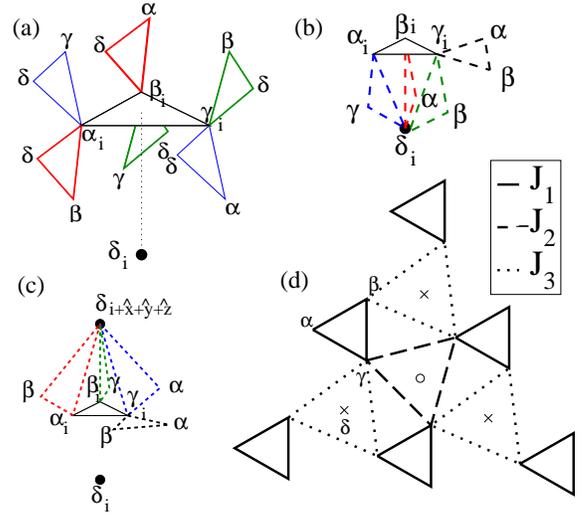}
\caption{\label{figure1} The 3D corner-shared equilateral triangle (trillium{\cite{trill}}) lattice. (a) Nearest neighbor (nn) bonds ($J_1$) form equilateral triangles with normal vectors along relative (1,1,1) directions, C$_3$ symmetry axes of the structure.  Each point (labeled $\alpha,\beta,\gamma,\delta$) belongs to three triangles.  The subscript $i$ denotes the elements of the cubic unit cell as given before Eq \ref{equation1}. (b) Second and (c) third nn bonds again form trillium lattices with normal vectors along the relative (1,1,1) directions. (d) Looking down the (1,1,1) axis, the 1st, 2nd and 3rd nn bonds and their relative distances and orientations are shown.  ``x'' marks the location of the nn $\delta$ sites one layer down and ``o'' one layer up.  Although the ground state of AF couplings on a trillium lattice features 120$^0$ rotated spins, it is not possible to simultaneously satisfy any two of $\{J_1,J_2,J_3\}$.      
}
\end{figure}

Recently, to describe the partially ordered state, the addition of a pseudoscalar order parameter was argued{\cite{Tewari}} to lead to a ``blue quantum fog'' with properties of a chiral liquid.  Attempts to explain the observed shift of weight from the (111) direction of the ambient pressure ordered helimagnet to the (110) direction at high pressures has led to theories which: i) favor a weakly ordered state resulting from introducing (non-local) mode-mode coupling interactions between helimagnetic structures to generate a new type of spin crystal{\cite{Binz}}; or ii) favor a lattice of cylindrical skyrmionic tubes with axes pinned to the (111) direction{\cite{bogdanov}}.

\begin{figure*}
\includegraphics[scale=0.24]{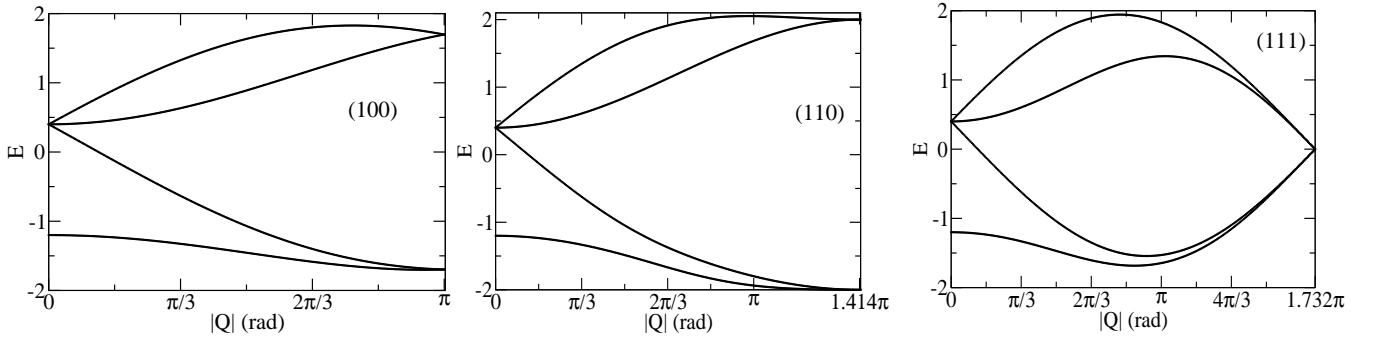}
\caption{\label{addon} Dispersion curves for $J_2/|J_1|=J_3/|J_1|=1/3$ and $|J_1|+J_2+J_3=1$ along the (100), (110) and (111) directions.  For all curves there is a mirror symmetry (not shown) about the left and right axes and we set $a_0$=1.  The ground state has energy -2$(|J_1|+J_2+J_3)$ with ${\bf{Q}}=(\pi,\pi,0)$ in agreement with the rotor model.}
\end{figure*}

However, the similarities between the MnSi lattice structure (B20) and the corner-shared tetrahedral (LiV$_2$O$_4$, (Y$_{0.97}$Sc$_{0.03}$)Mn$_2$) and distorted windmill ($\beta$-Mn) lattices have been overlooked so far.  Heisenberg models have been used to successfully explain anomalous behavior in these itinerant systems{\cite{liv2o4,yscmn,betamn}}.  Moreover, the same anomalous resistivity $\propto T^{\frac{3}{2}}$ near and above the pressure tuned quantum critical point has been reported in FeGe{\cite{pedra}}, which has the same B20 lattice structure.  This raises the possibility that the lattice structure may play a role in understanding the helical magnetic orderings in MnSi.  In this paper, we suggest a microscopic model consisting of Heisenberg interactions on the MnSi lattice.  We find that a spiral ordering along (110) is naturally realized as a result of a competition between FM and antiferromagnetic (AFM) interactions.

{\it{Model:}} The unit cell of MnSi is a simple cubic structure, and contains four atoms at sites: $\{\delta_i,\alpha_i,\gamma_i,\beta_i\}=\{(u,u,u);(u+\frac{1}{2},\frac{1}{2}-u,1-u);(1-u,u+\frac{1}{2},\frac{1}{2}-u);(\frac{1}{2}-u,1-u,u+\frac{1}{2},)\}a_0$ with $u_{Mn}=0.138$ and $a_0=4.56\AA$\cite{ishikawa}.  Si atoms are located in the same pattern with $u_{Si}=0.854$.  Mn atoms form corner-sharing equilateral triangular (hereafter trillium{\cite{trill}}) lattices as shown in Fig. \ref{figure1}.  Each Mn atom is shared by 3 triangles and the normal vectors to the triangles lie along the 4 body diagonals of the cubic unit cell: (1,1,1), (1,-1,-1), (-1,-1,1) and (-1,1,-1).  We study a classical Heisenberg model with nearest neighbor (nn) FM interactions ($J_1<0$) and next- ($J_2>0$) and third- ($J_3>0$) nn AFM couplings on the 3 dimensional (3D) trillium lattices formed by the Mn sites of MnSi,
\begin{eqnarray}
H&=&J_1\sum_{\langle ij\rangle}\overrightarrow{s}_i\cdot \overrightarrow{s}_j+J_2\sum_{\langle\langle ij\rangle\rangle}\overrightarrow{s}_i\cdot \overrightarrow{s}_j+J_3\sum_{\langle\langle\langle ij\rangle\rangle\rangle}\overrightarrow{s}_i\cdot \overrightarrow{s}_j\nonumber\\&=&\sum_{\zeta=1}^3H^{(\zeta)}=\sum_{\zeta=1}^3\sum_Q\overrightarrow{S}_Q(2J_{\zeta}\overleftrightarrow{M}_Q^{(\zeta)})\overrightarrow{S}_{-Q}.\label{equation1}
\end{eqnarray}
Here $J_{\zeta}$ represents the $\zeta$th order Heisenberg coupling with $\zeta$=1-3 and $\overrightarrow{S}_Q=(\overrightarrow{s}^{\delta}_Q,\overrightarrow{s}^{\alpha}_Q,\overrightarrow{s}^{\gamma}_Q,\overrightarrow{s}^{\beta}_Q)$ where $\overrightarrow{s}^{\nu}_i=\sum_Q \overrightarrow{s}^{\nu}_Q e^{iQ r^{\nu}_i}$.  $\overleftrightarrow{M}_Q^{(\zeta)}$ is the matrix given by,
\begin{widetext}
\begin{eqnarray}
\left(\begin{array}{cccc}0&\cos(\eta q_xa_0)e^{i(\mu q_ya_0 + \nu q_za_0)}&\cos(\eta q_ya_0)e^{i(\nu q_xa_0+\mu q_za_0 )}&\cos(\eta q_za_0)e^{i(\mu q_xa_0+\nu q_ya_0)}\\\cos(\eta q_xa_0)e^{-i(\mu q_ya_0 + \nu q_za_0)}&0&\cos(\eta q_za_0)e^{i(\mu q_xa_0 - \nu q_ya_0)}&\cos(\eta q_ya_0)e^{i(\nu q_xa_0-\mu q_za_0 )}\\\cos(\eta q_ya_0)e^{-i(\nu q_xa_0+\mu q_za_0 )}&\cos(\eta q_za_0)e^{-i(\mu q_xa_0 - \nu q_ya_0)}&0&\cos(\eta q_xa_0)e^{i(\mu q_ya_0 - \nu q_za_0)}\\\cos(\eta q_za_0)e^{-i(\mu q_xa_0+\nu q_ya_0)}&\cos(\eta q_ya_0)e^{-i(\nu q_xa_0-\mu q_za) )}&\cos(\eta q_xa_0)e^{-i(\mu q_ya_0 - \nu q_za_0)}&0\end{array}\right).
\label{equation5}
\end{eqnarray}
\end{widetext}
where $\eta$=$\frac{(2n+1)}{2}$, $\mu$=$(2u+m-\frac{1}{2})$, $\nu$=$(2u+p)$ and the three integers $\{m,n,p\}$ take the values $\{0,0,0\},\{0,0,-1\}$ and $\{1,0,0\}$ for $\zeta = 1,2$ and 3 respectively.  

A mean field (MF) treatment which corresponds to the diagonalization of the matrix of $\overleftrightarrow{M}_Q^{(1)}$ has been studied{\cite{earlier}}.   The ground state of $H^{(1)}$ with energy -3$J_1$ shows degenerate states satisfying the equation, 
\begin{eqnarray}
\cos^2(\eta q_x a_0)+\cos^2(\eta q_y a_0)+\cos^2(\eta q_z a_0)=\frac{9}{4}.\label{equation4}
\end{eqnarray}
 Here $\overrightarrow{Q}(=(q_x,q_y,q_z))$ is a wavevector of the first Brillouin zone and the MF ground state is given by the minimum of the eigenvalues $\lambda_Q^{\xi}$ at $E=\lambda_{Q_0}^{\xi}$ where $\xi$ represents 4 different eigenvalues, and $Q_0$ is where the minimum occurs in $Q$. The set of wavevectors satisfying Eq. \ref{equation4} form a sphere-like shape which resembles the sphere of wavevectors observed in MnSi, while the peculiar anisotropic intensities cannot be captured.  However, a Monte Carlo treatment{\cite{sergeius}} on this lattice found an ordered state in agreement with the result obtained by a classical rigid rotor model{\cite{earlier}}.  The ground state of  $H^{(1)}$ forms a lattice of 120$^0$ rotated spins on each triangle with the wavevector along the (100) direction. It is interesting to note that, unlike on the trillium lattice, MF and Monte Carlo results agree on the corner-shared tetrahedral lattice, common to the pyrochlore and spinel structures..

It is straightforward to see that the solely 2nd (or 3rd) nn Heisenberg model generically has the same structure\cite{note} and ground state as the 1st nn model.  However, as indicated in Fig. \ref{figure1}, it is not possible to simultaneously satisfy AFM couplings for any two of these lattices.  Below we will show that the microscopic model of Eq. \ref{equation1} finds a ground state with spiral ordering along (110) due to the competition between AFM couplings and between AFM and FM couplings.  We use both MF and the rigid rotor approximations to avoid any artificial effects of the MF approximation as found in the nearest neighbor AFM model.  As discussed elsewhere{\cite{earlier}}, the rigid rotor approximation corresponds to the mapping of a spin to a classical vector with 2 independent angles.  Minimizations of energy with respect to configurations of these real space rigid rotors allow one access to the physical spin structure of the ground state.

\begin{figure}
\includegraphics[scale=0.3]{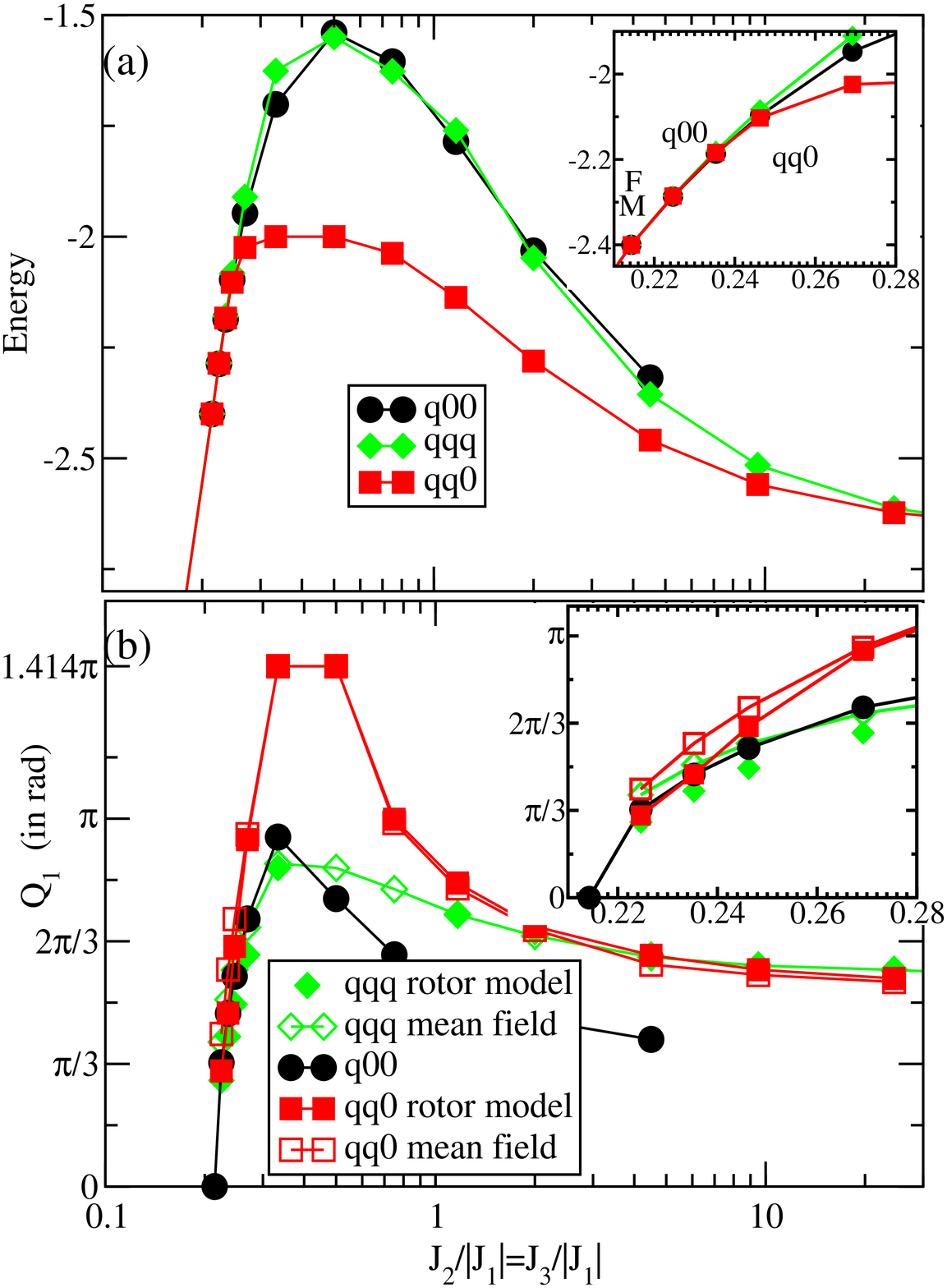}
\caption{\label{figure3} For $J_1 < 0$, $J_2= J_3>0$, and $|J_1|+J_2+J_3$ = 1 and three different directions $(100)$, $(110)$ and $(111)$: (a) Energy (in units of ($|J_1|+J_2+J_3$)) versus $J_2/|J_1|$.  For $J_2/|J_1| \lesssim 0.214$ the ground state is FM with energy $6(J_1+J_2+J_3)$.  For $J_2/|J_1|>0.214$, the ground state orders along $(110)$, except for tiny regions of $(100)$ ordering very close to FM (see inset), and $(111)$ order when $J_2/|J_1|\rightarrow\infty$. (b) Magnitude of the ordering wavevector, $Q_1$, at the minimal energy solution of a given direction versus $J_2/|J_1|$.  Notice that the minima have similar $Q_1$ magnitudes near the onset of FM.  Mean field theory tends to over/(under)estimate $Q_1$ for FM/(AFM) dominated couplings for the $(110)$ and $(111)$ symmetries.  For the $(110)$ ordered ground state, a small component of the spins (orthogonal to the component rotating with $Q_1$) at the $\{\alpha,\gamma\}$/($\{\delta,\beta\}$) sites retains a $(\pi/a_0,\pi/a_0,0)$ symmetry. 
}
\end{figure}

{\it{Phase diagram:}} We studied the energy dispersion curves in $\overrightarrow{Q}$ within the MF approximation for a given $J_2/|J_1|$, and we set $J_2=J_3$ for all cases.  We found the ground state to be FM for $J_2/|J_1|\lesssim 0.214$.  As $J_2(=J_3)$ increases, spiral ordering with a finite wave vector becomes the ground state.  The energy dispersion curves for three different directions, (100), (110), and (111), are shown in Fig. \ref{addon} for $J_2/|J_1|=1/3$.  As shown, we found a ground state with energy -2$(|J_1|+J_2+J_3)$ along the (110) direction with magnitude $Q_1=|\overrightarrow{Q}|$ of $\sqrt{2}\pi/a_0$ rad.  The rotor model minimizes the energy along the (110) direction for the same value of $J_2/|J_1|$ (=1/3) verifying the MF result.  We present the ground state energies of the rotor model as a function of $J_2/|J_1|$ for three different directions in Fig. \ref{figure3} (a).  It is clearly shown that magnetic ordering along (110) has a lower energy than along the (100) and (111) directions for virtually the entire phase diagram with exceptions close to $J_2^{cri} (=0.214 |J_1|)$ and $J_2/|J_1|\rightarrow\infty$.  Close to $J_2^{cri}$, various states represented by different directions of the ordering wavevectors are almost degenerate.  We also present the magnitude of the ordering wavevector $Q_1=|\overrightarrow{Q}|$ for the corresponding states in Fig. \ref{figure3} (b).  The MF and rotor model approximations are found to agree for 0.32$\lesssim J_2/|J_1|\lesssim0.5$.  However, the MF approximation overestimates the value of $Q_1$ compared to the rotor model for $J_2^{cri}\lesssim J_2 < 0.32 |J_1|$, while it underestimates the value of $Q_1$ for 0.5$|J_1|< J_2$.  It is also interesting to note that the size of $Q_1$ for the (110) direction is generally larger than those of the other two directions.

\begin{figure}
\includegraphics[scale=0.35]{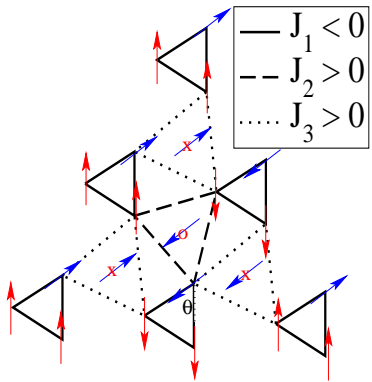}
\caption{\label{figure2}Looking down the (111) axis.  At the special point $J_1=-2J_2=-2J_3$, the ground state spin structure has (110) symmetry and spin arrangements form two independent sets of pairs $\{\alpha,\gamma\}$ and $\{\delta,\beta\}$ (ie. the relative angle, $\theta$, between these two sets is completely arbitrary.) Two spins on each FM ($J_1$) bonded triangle are parallel, while two spins on each AFM ($J_2$ and $J_3$) bonded triangle are anti-parallel.  An analogous state is not available to (100) and (111) symmetry spin structures.}
\end{figure}

{\it{Spin structures:}} As we discussed, an advantage of the rotor model is that we can identify the spin structure of a given state.  Spin structures periodic along (110) tend to break the four-site unit cell into two sublattices. Indeed, all ground state spin structures with this symmetry feature parallel spins on the nn $\beta_i$ and $\delta_{i+\hat{y}}$ sites (here $\hat{y}$ denotes a coordinate increase by one unit in the $y$-direction). To illustrate this, let us consider the special point, $J_1=-J_2-J_3$, $J_2=J_3>0$, where a further choice of parallel spins at the $\gamma_i$ and $\alpha_i$ sites minimizes the energy  at $-2(|J_1|+J_2+J_3)$ with    
 wavevector $(\frac{\pi}{a_0},\frac{\pi}{a_0},0)$.
In this configuration, two spins on each FM ($J_1$) bonded triangle are parallel, while two spins on each AFM ($J_2$ and $J_3$) bonded triangle are anti-parallel as shown Fig. \ref{figure2}.  The energy is independent of the relative orientation of the third spin on each triangle, as FM and AFM interactions cancel, leaving the structure independent of an overall relative angle between pairs of spins.

 For a coplanar spiral spin structure (even one with structure inside the unit cell as here), all spin components should lie in a plane which, for convenience, here we will define to be perpendicular to the wavevector $\overrightarrow{Q}_1$.  This is the case for $0.32\lesssim J_2/|J_1| \lesssim 0.5$ where the MF and rotor models agree.  For a 3D spin structure, a non-coplanar spin structure may be energetically favorable, where one or two spins of the four site unit cell slightly tilt out of this plane.  Further, the spin component out of the plane can alternate in sign, exhibiting a distinct wavevector, $\overrightarrow{Q}_2$.  The two sublattice nature of the spin structure continues away from the special point at $J_2/|J_1|(=J_3/|J_1|)=1/2$.  For $J_2/|J_1|\lesssim 0.32$, spins at the $\alpha$ and $\gamma$ sites acquire a component parallel to $\overrightarrow{Q}_1$ with wavevector $\overrightarrow{Q}_2 = (\pi/a_0,\pi/a_0,0)\ne\overrightarrow{Q}_1$.  For $J_2/|J_1|>0.5$ a similar noncoplanar spin structure arises but at the $\delta$ and $\beta$ sites.  In the calculation of the neutron scattering structure factor no interference terms arise between the spin components with wavevectors $\overrightarrow{Q}_1$ and $\overrightarrow{Q}_2$, as the corresponding spin structures are orthogonal.  The only effects of the development of this second ordering wavevector on the neutron scattering structure factor would be to slightly decrease the weight at $\overrightarrow{Q}_1$, and to create a tiny peak at $\overrightarrow{Q}_2 =(\pi/a_0,\pi/a_0,0)$.  We have extracted the components of the spin structure along and perpendicular to $\overrightarrow{Q}_1$ from the ground state (110) spin structures as a function of $J_2/|J_1|$ although we omit this here for brevity.

{\it{Discussion:}} It is interesting to ask whether this $(110)$ ordered state might have any relevance to the weakly ordered $(110)$ state found by Pfleiderer et al{\cite{pflei}}. We know from measurements\cite{pflei2} of magnetic susceptibility, ($\chi$), that the low temperature Curie law maintains its slope as pressure increases past the transition to the anomalous NFL phase. The intercept of $\chi^{-1}$ appears to eventually reach a weakly negative offset consistent with the presence of some AFM couplings.  Such couplings might arise in an RKKY-based picture of itinerant magnetism between local moments on the Mn sites.  The pitch of the structured helimagnet found here is an order of magnitude larger than that observed in MnSi.  This can be explained by the inclusion of further neighbor couplings which also form trillium lattices.  We expect that in this way an analogous ordering along (110) can be obtained with a smaller wavevector, although the derivation of the strength and size of such couplings remains to be investigated.

 At ambient pressure, a good approximation to the ground state of MnSi is given by a simple spin spiral with $(111)$ symmetry. This symmetry pins at low energies from a nearly degenerate set of fixed period simple spirals sampled at higher temperatures.  The lack of inversion symmetry of the MnSi lattice permits the introduction of asymmetric spin terms arising from the spin-orbit coupling on the lattice.   The addition of DM-vectors parallel to the nearest neighbor Mn-Mn bonds on this lattice nicely reproduces a nearly degenerate surface in momentum space as will be presented elsewhere{\cite{us}}. With the noted exception of the FM phase (which becomes a $(100)$ ordered phase), we find that the phase diagram of Fig. \ref{figure3} is very weakly effected by competition from these first-neighbor DM-vectors.  

 Briefly we comment on how the $(111)$ ordered ambient pressure phase is thought to pin from the degenerate helimagnetic phase.  The local environment of the Mn site is surrounded by 7 Si atoms, with the shortest Mn-Si distance along the local (1,1,1) axis.  It therefore seems plausible that crystal/ligand field effects\cite{andersen} might produce a local (1,1,1) anisotropy of the spin structure.  One term which might arise from a full consideration of spin-orbit interactions is of the form, $-\zeta \sum_{i,\nu}(\overrightarrow{s}_{i,Ising}^{\nu})^4$.  Here the sum runs over the number of unit cells and the 4 sites within the unit cell, $\overrightarrow{s}_{i,Ising}^{\nu}=(\overrightarrow{s}_i\cdot\overrightarrow e_{\nu})\overrightarrow e_{\nu}$ and $\overrightarrow e_{\nu}$ is the unit vector associated with the local (1,1,1) anisotropy at site $\nu$.  Assuming this term is small, we can evaluate the energy change that such a term would introduce to the simple helimagnet.  For $\zeta>0$ this term is found to lower the energy of the (100), (110) and (111) directions by -6$\zeta$, -6.5$\zeta$ and -8$\zeta$ respectively per unit cell, effectively favoring (1,1,1).  A pressure increase likely increases the overlap of the various Si atoms with the Mn sites (as apparently seen in FeGe\cite{pedra}), reducing the strength of pinning.

{\it{Summary:}} In this paper we have provided a simple classical spin model on the MnSi lattice which exhibits an extended phase with a $(110)$ ordered ground state.  We have considered an extended Heisenberg model with both FM and AFM couplings joining the Mn sites.  Within this microscopic model, we find that the corner-shared equilateral triangle structure of the lattice plays a crucial role in the suppression of both $(100)$ and $(111)$ orderings.    Even though AFM couplings are known to favor $(100)$ order{\cite{us}}, competitions between AFM $J_2$ and $J_3$ frustrate this order and are found to yield $(110)$ order in the presence of a FM $J_1$, with or without a nn DM interaction. Such physics has not been considered in LG studies.  A detailed derivation of the RKKY interaction on this unusual lattice is desired and could be anticipated to lead to a weak $(110)$ ordering arising from a competition between AFM and DM interactions.  It is tempting to speculate that the crossover from $(111)$ to weak $(110)$ order seen in MnSi might then result from the application of pressure reducing crystal field splittings and increasing AFM coupling strengths.

{\it{Acknowledgments:}} This work was supported by an NSERC PDF, an NSERC CRC, an Alfred P. Sloan Fellowship, NSERC and the CIAR.  




\begin{thebibliography}{00}
\bibitem{trill} We named this lattice in Ref.\cite{earlier} to distinguish it from the previously studied distorted windmill lattice of $\beta$-Mn which also features a 3D lattice of corner-shared equilateral triangles.
\end{thebibliography}
\end{document}